\documentclass[aps,prl,reprint,showpacs]{revtex4-1}

\usepackage{graphicx}
\usepackage{amsfonts}
\usepackage{amsmath}
\usepackage{color}
\usepackage[hypertexnames=false]{hyperref}

\begin{document}

\title{Symmetric States Requiring System Asymmetry}

\author{Takashi Nishikawa} 
\email[Correspondence and requests for materials should be addressed to T.N. (t-nishikawa@northwestern.edu)]{}
\author{Adilson E. Motter}
\affiliation{Department of Physics and Astronomy, Northwestern University, Evanston, IL 60208, USA}
\affiliation{Northwestern Institute on Complex Systems, Northwestern University, Evanston, IL 60208, USA}

\begin{abstract}
Spontaneous synchronization has long served as a paradigm for behavioral uniformity that can emerge from interactions in complex systems. When the interacting entities are identical and their coupling patterns are also identical, the complete synchronization of the entire network is the state inheriting the system symmetry. As in other systems subject to symmetry breaking, such symmetric states are not always stable.  Here we report on the discovery of the converse of symmetry breaking---the scenario in which complete synchronization is not stable for identically-coupled identical oscillators but becomes stable when, and only when, the oscillator parameters are judiciously tuned to nonidentical values, thereby breaking the system symmetry to preserve the state symmetry. Aside from demonstrating that diversity can facilitate and even be required for uniformity and consensus, this suggests a mechanism for convergent forms of pattern formation in which initially asymmetric patterns evolve into symmetric ones. 
\end{abstract} 

\href{http://dx.doi.org/10.1103/PhysRevLett.117.114101}{Phys. Rev. Lett. {\bf 117}, 114101 (2016)}\\

\vspace{3mm}

\maketitle

Symmetry---the property of appearing the same from different viewpoints---is so central to physics that Weyl~\cite{Weyl1952} suggested that ``all a priori statements in physics have their origin in symmetry'';
Anderson~\cite{Anderson1972}  went further to propose that ``physics is the study of symmetry.''
In the study of complex networks this tradition was for many years relegated to a secondary position, for the excellent reason that real complex systems appeared not to exhibit symmetries.
Recent work has shown, however, that they not only can exhibit a myriad of symmetries~\cite{MacArthur2008} but also that such symmetries have direct implications for dynamical behavior (see Ref.~\cite{Pecora2014} for example). 
Partially motivated by that, significant recent attention has been dedicated to the extreme, most symmetric case of uniform networks in which nodes are all identically coupled to the others and have no natural grouping, as in a ring or all-to-all network.
It has been shown that such systems can exhibit spatiotemporal patterns of coexisting synchronous and non-synchronous behavior~\cite{Kuramoto2002,Abrams2004}, for which elaborated mathematical analysis techniques are now available~\cite{Ott2008}.
The emergence of these patterns can be regarded as a form of symmetry breaking, since the realized state has less symmetry than the system~\cite{Motter2010}.
Here we demonstrate for the first time that the converse of symmetry breaking with the roles of the system and its state reversed---which we term {\it asymmetry-induced symmetry}---is also possible. 
We provide examples of uniform, rotationally symmetric networks of coupled oscillators for which stable uniform states (thus rotationally symmetric states) do not exist when the nodes are identical but do exist when the nodes are not identical.

In a network of coupled oscillators a uniform, symmetric state represents synchronization, in which all units swing in concert, following the exact same dynamics as a function of time~\cite{Pikovsky2003}.
Synchronization dynamics is widespread across fields---ranging from physics and engineering to biology and  social sciences---and is intimately related to the twin processes of consensus dynamics and convergence to uniform patterns.
Consensus dynamics is a process through which a network of interacting agents can achieve a common objective or reach agreement.
Examples include decentralized coordination of moving sensors~\cite{4140748,4700861} and the dynamics of collective opinion formation in social networks~\cite{DeGroot:1974,Friedkin:2011}.
Convergence to uniformity can occur through processes of diffusion or relaxation, in which pairwise interactions in the network tend to reduce the difference between the states of the nodes.
Examples of such processes include convergence to equilibrium in chaotic chemical reaction systems~\cite{Showalter:2015,Tompkins:2015}, population dispersion in natural systems~\cite{Nakao:2010fk}, and relaxation in fluid networks~\cite{maas1987transportation}.

\begin{figure*}
\includegraphics[width=5.5in]{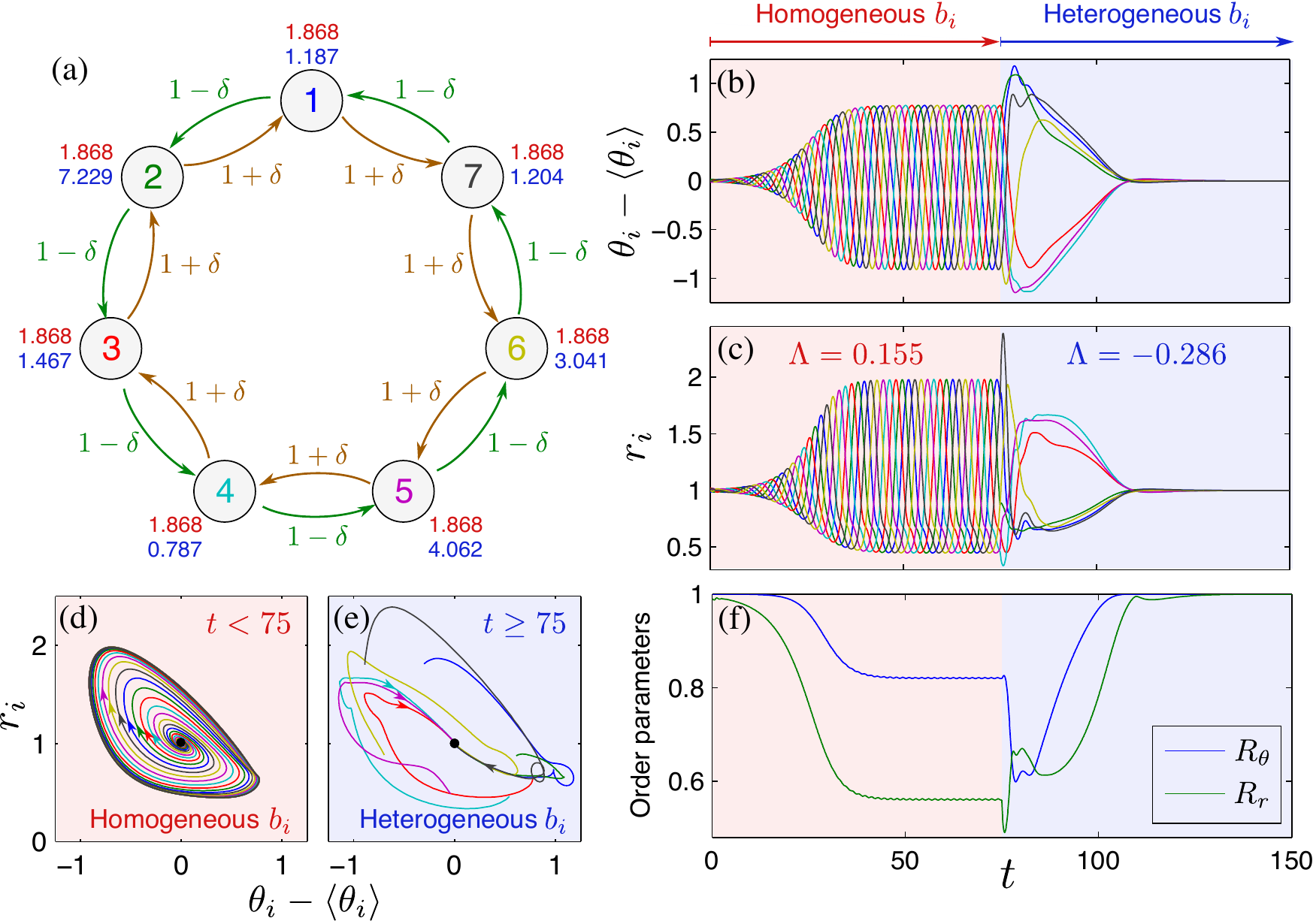}
\caption{
Oscillator heterogeneity stabilizes homogeneous synchronous state in homogeneous network.
(a) Homogeneous network of $n=7$ nodes.
The red (top) and blue (bottom) numbers are the oscillators' $b_i$ values used for $t < 75$ and $t \ge 75$, respectively, in our simulation of Eq.~\eqref{eqn:system} for $\varepsilon = 2$ and $\delta=0.3$.
(b--e) Oscillator state trajectory showing desynchronization with homogeneous $b_i=b^*$ when $t<75$, followed by spontaneous synchronization with heterogeneous $b_i$ when $t \ge 75$.
(b) phase angle $\theta_i$ (relative to their average $\langle\theta_i\rangle$) vs.\ $t$. 
(c) amplitude $r_i$ vs.\ $t$.
(d,e) $r_i$ vs.\ $\theta_i - \langle\theta_i\rangle$ for $t<75$ (d) and $t\ge 75$ (e).
The synchronous state corresponds to $\theta_i - \langle\theta_i\rangle = 0$, $r_i = 1$.
(f) Order parameters $R_{\theta}$ and $R_r$ quantifying the degree of synchronization. They are defined by $R_{\theta} := \frac{1}{n}\bigl\vert \sum_{i} \exp(\mathbf{i}\theta_i) \bigr\vert$ and $R_r := \exp(-\sigma_r)$, respectively, where $\mathbf{i} := \sqrt{-1}$ is the imaginary unit and the standard deviation $\sigma_r$ is computed as $\sigma_r^2 := \frac{1}{n-1}\sum_i (r_i - \langle r_i \rangle)^2$. See \cite{methods} for details and an animation of the dynamics.}
\label{fig1}
\end{figure*}

As a model system that can exhibit asymmetry-induced symmetry, we introduce a network of $n$ two-dimensional oscillators whose dynamics is governed by
\begin{equation}\label{eqn:system}
\begin{split}
\dot{\theta}_i &= \omega + r_i - 1 - \gamma r_i \sum_{j=1}^n\sin(\theta_j - \theta_i),\\
\dot{r}_i &= b_i r_i ( 1 - r_i) + \varepsilon r_i \sum_{j=1}^n A_{ij} \sin(\theta_j - \theta_i),
\end{split}
\end{equation}
where $\theta_i$ and $r_i$ are the angle and amplitude variables for the $i$th oscillator, respectively, the constants $\omega$ and $b_i>0$ characterize the dynamics of individual oscillators, the parameters $\gamma>0$ and $\varepsilon > 0$ are constants representing the overall coupling strength, and $\mathbf{A} = (A_{ij})_{1 \le i,j \le n}$, $A_{ij} \geq 0$, is the adjacency matrix encoding the structure of the (possibly weighted and directed) network. 
Note that the interaction network of system~\eqref{eqn:system} has two components, one representing the uniform, angle-to-angle coupling between all pairs of nodes, and the other representing the angle-to-amplitude coupling with the network structure given by the matrix $\mathbf{A}$.
For arbitrary network structure $\mathbf{A}$, the system~\eqref{eqn:system} has a synchronous state given by
\begin{equation}\label{eqn:sync-state}
\theta_1(t) = \cdots = \theta_n(t) \equiv \theta_0 + \omega t, \quad r_1(t) = \cdots = r_n(t) \equiv 1,
\end{equation}
in which each oscillator follows the limit cycle of the isolated oscillator dynamics~\cite{methods}.
This state is guaranteed to exist because all the coupling terms vanish when $\theta_1 = \cdots = \theta_n$.  We see from the form of Eq.~\eqref{eqn:system} that the coupling between the angle and amplitude variables tends to stabilize the synchronous state, while the coupling within the angle variables (through all-to-all topology and a negative coupling strength, $-\gamma < 0$) tends to destabilize it.  The balance between the two effects determines the synchronization stability, which can be quantified by the maximum Lyapunov exponent $\Lambda$, defined as the exponential rate of convergence to (if $\Lambda < 0$) or divergence from (if $\Lambda > 0$) the synchronous state (see \cite{methods} for details on the stability analysis).
We consider the class of uniform networks in which nodes are arranged in a one-dimensional ring and each node is identically coupled to the rest of the network.
Specifically, for a given parameter $\delta$, each node $i$ receives input from node $i-1$ with coupling strength $1-\delta$ and from node $i+1$ with strength $1+\delta$ (where we have defined the indices $i=0$ and $i=n+1$ to denote the nodes $i=n$ and $i=1$, respectively).
An example of such a network is illustrated in Fig.~\ref{fig1}(a) for $n=7$.
Here we assumed that the average coupling strength is one for the two links pointing to each node, but the more general class of networks for which this average is arbitrary can be reduced to the class we have just defined by factoring out a scalar from $A_{ij}$ and having it absorbed into the parameter $\varepsilon$ in Eq.~\eqref{eqn:system}.
Model~\eqref{eqn:system} represents a wide range of other systems that can exhibit asymmetry-induced symmetry.  For example, a general class of networks of coupled Stuart-Landau oscillators~\cite{Daido:2004,Bordyugov:2010,Laing:2010,Zakharova:2014,Panaggio:2015} (whose node dynamics is based on the normal form for an oscillator near a supercritical Hopf bifurcation~\cite{kuramoto2003chemical}) can be parametrized in such a way that the parametric dependence of synchronization stability is identical to that for model~\eqref{eqn:system}~\cite{methods}. 

\begin{figure*}
{\includegraphics[width=5.5in]{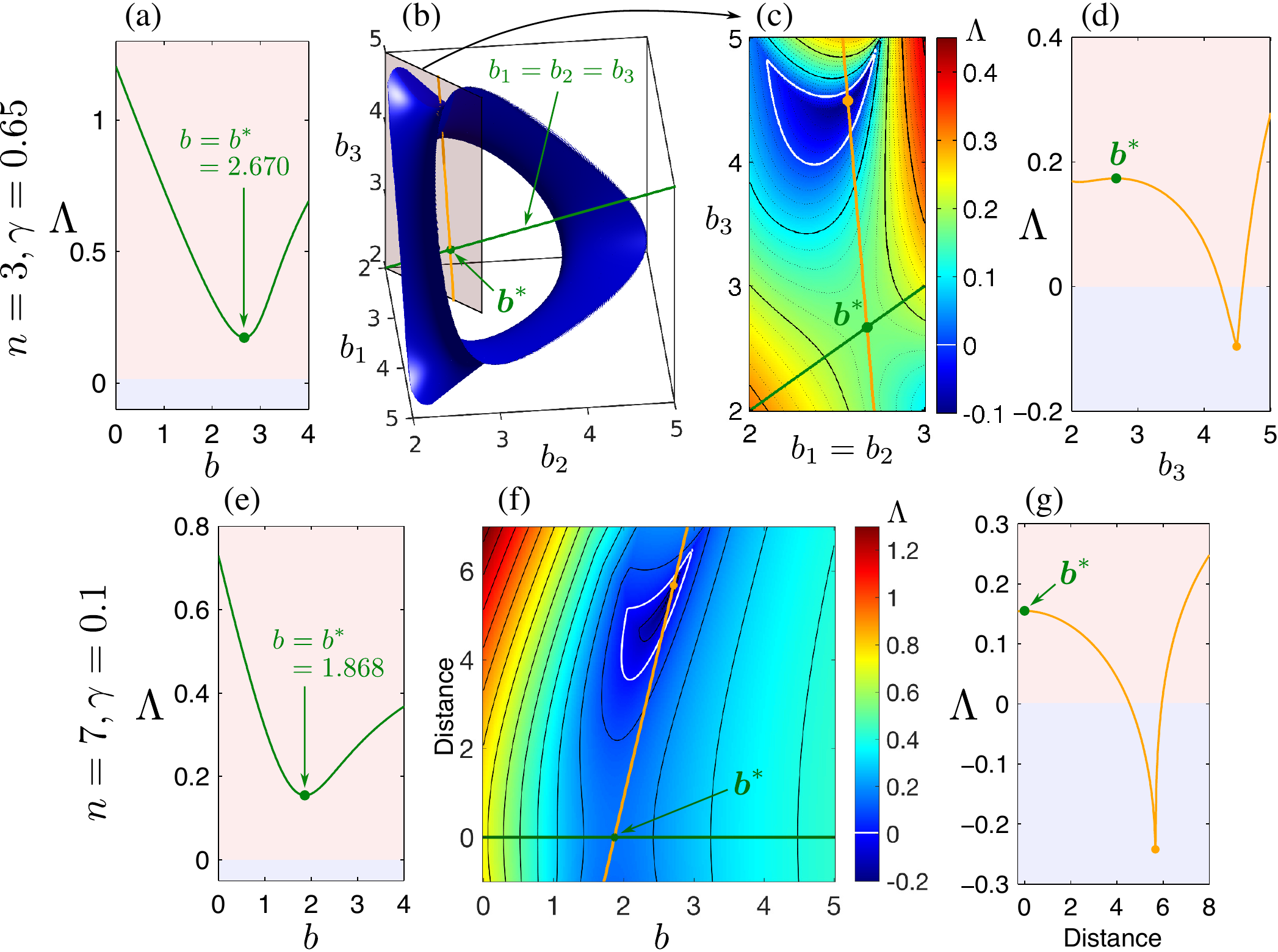}}
\caption{
Stability landscape for the synchronous state.
(a--d) Maximum Lyapunov exponent $\Lambda$ for $n=3$. 
(a) $\Lambda$ vs.\ $b$ for $b_i = b$, $\forall i$.
(b) Region of stability, $\Lambda(b_1,b_2,b_3) \le 0$ (blue) in the full $\boldsymbol{b}$-space.
(c) $\Lambda$ on the slice shown in (b).
(d) $\Lambda$ vs.\ $b_3$ along the orange line in (b) and (c).
(e,f) $\Lambda$-landscape for the $n=7$ case in Fig.~\ref{fig1}.
(e) $\Lambda$ vs.\ $b$ for $b_i = b$, $\forall i$.
(f) $\Lambda$ on a 2D slice of the 7D $\boldsymbol{b}$-space, parametrized by $b$ along the line $b_i = b$ and the (Euclidean) distance from that line.
The slice was selected to contain the (orange) point corresponding to the heterogeneous $b_i$ used in Fig.~\ref{fig1}.
The white curves in (c) and (f) indicate $\Lambda=0$.
(g) $\Lambda$ vs.\ the distance along the orange line in (f).
We used $\varepsilon = 2$ and $\delta=0.3$ for all panels.}
\label{fig2}
\end{figure*}

Figure~\ref{fig1} shows the dynamics demonstrating asymmetry-induced symmetry for the example system~\eqref{eqn:system}.  
For identical values of $b_i$, which make the oscillators identical, the synchronous state is unstable, even when the common value is chosen to be the one that minimizes $\Lambda$, which we denote by $b^*$.
In this case the system starting near the synchronous state diverges away and approaches a traveling wave state (see \cite{methods} for an animation of this state). 
However, if we allow for nonidentical values of $b_i$, we can stabilize the synchronous  state.
Indeed, as shown in Fig.~\ref{fig1}, after switching to a numerically identified combination of non-homogeneous $b_i$ values, we see that the oscillators spontaneously return to the synchronous state.
Thus, for system~\eqref{eqn:system}, 
{\it the stability of the (uniform) synchronous state can only be supported by nonidentical oscillators.}
While we focus here on uniform networks to avoid confounding factors (e.g., differences between oscillators needed to compensate for differences between their couplings), the conclusion that inherent heterogeneity can be necessary to realize uniform states is general and also valid for nonuniform networks (see \cite{methods} for concrete examples).

The landscape of stability in the space of all possible $\boldsymbol{b} := (b_1,\ldots,b_n)$ provides a more complete view of asymmetry-induced symmetry.
Along the diagonal line $b_1=\cdots=b_n \equiv b$ in this space, $\Lambda$ as a function of $b$ typically has a single minimum at $b = b^*$ with $\Lambda(b^*) > 0$, in which case no homogeneous oscillators can be stably synchronized in the form of Eq.~\eqref{eqn:sync-state}.
Figures~\ref{fig2}(a) and \ref{fig2}(e) show example cases for $n=3$ and $n=7$, respectively, in which $\Lambda(b^*) > 0$ (which is satisfied even when considering both positive and negative $b$).
In the full $n$-dimensional $\boldsymbol{b}$-space, however, there can be a significant (nonzero-volume) region of stable synchronization [see Fig.~\ref{fig2}(b) and Fig.~\ref{fig2}(f)].
The shape of this region is necessarily cyclically symmetric around the homogeneous-$b_i$ line due to the symmetry of the network dynamics with respect to cyclic permutations of the nodes.
This can be seen in the case of $n=3$, shown in Fig.~\ref{fig2}(b), in which the stability region (blue) is invariant under the $120^\circ$ rotation around that line.
In both $n=3$ and $n=7$ cases, we observe that the stability region lies far away from the diagonal line representing the homogeneous-oscillator networks [the green lines in Fig.~\ref{fig2}(b) and Fig.~\ref{fig2}(f)], indicating that significant differences between the oscillators are required to achieve stable synchronization.
For $n=3$, the stability region also appears to have a mirror symmetry about the three planes $b_1 = b_2$, $b_2 = b_3$, and $b_3 = b_1$.
Associated with these planes we find six points of maximum stability in the box shown in Fig.~\ref{fig2}(b): three pairs related to each other by the $120^\circ$ rotation about the diagonal line, with each pair symmetrically located about and very close to one of the planes (at distance $\approx 0.010$).
The pair associated with the plane $b_1 = b_2$ is $\boldsymbol{b} \approx (2.560, 2.575, 4.495)^T$ and $(2.575, 2.560, 4.495)^T$.
Thus, despite the symmetry of the stability region, the individual points of maximum stability are not symmetric and correspond to having distinct parameters for the oscillators.

How does the shape of this stability landscape depend on the system parameters $\varepsilon$, $\gamma$, and $\delta$?  
It is sufficient to consider just $\gamma$ and $\delta$, since we can show~\cite{methods} that 
\begin{equation}\label{eqn:lambda-max-scaling}
\Lambda(b_1,\ldots,b_n; \varepsilon, \gamma, \delta)
= \bigl(\textstyle\frac{\varepsilon}{\varepsilon_0}\bigr)^{\frac{1}{2}} \cdot
\Lambda(b'_1, \ldots, b'_n; \varepsilon_0, \gamma', \delta),
\end{equation}
where $\gamma' := \gamma\sqrt{\varepsilon_0/\varepsilon}$ and $b'_i := b_i\sqrt{\varepsilon_0/\varepsilon}$.
Thus, the landscape for arbitrary $\varepsilon$ is identical to an $\varepsilon$-scaled version of the landscape for $\varepsilon=\varepsilon_0$ with $\gamma'$ and $b'_i$.
We therefore fix $\varepsilon = 2$ and discuss dependence on $\gamma$ and $\delta$ in the following.

\begin{figure}[t]
\begin{center}
\includegraphics[width=\columnwidth]{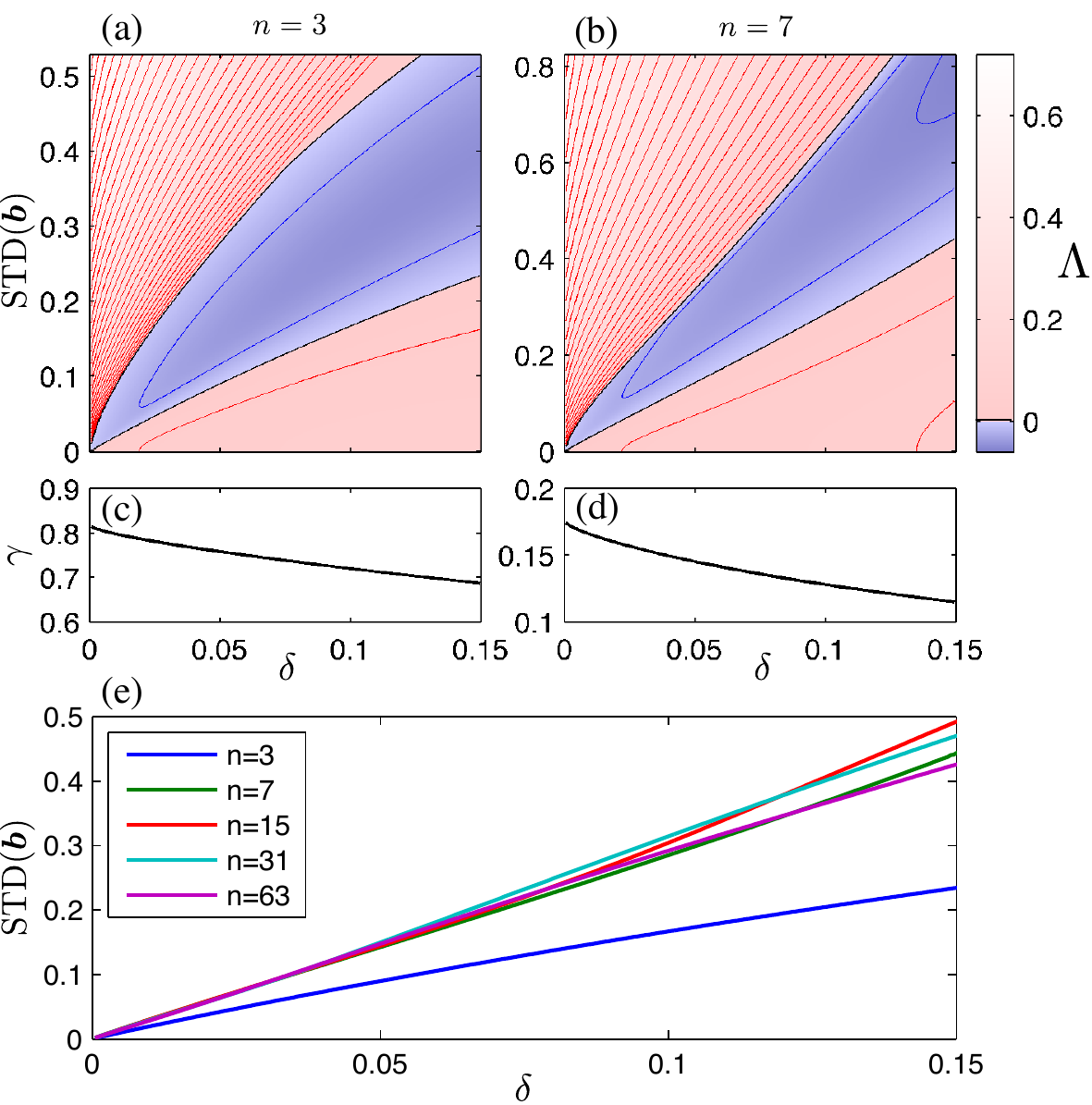}
\caption{
Network directionality enables asymmetry-induced symmetry.
For $\varepsilon = 2$ and a given $\delta$, quantifying the directionality, we identify a $\gamma$ value for which $\Lambda(\boldsymbol{b}^*) > 0$ but $\Lambda(\widetilde{\boldsymbol{b}}) < 0$ for some heterogeneous $\widetilde{\boldsymbol{b}} \neq \boldsymbol{b}^*$ (see \cite{methods} for details).
(a,b) Color-coded $\Lambda$ as a function of $\delta$ and the $b_i$-heterogeneity (measured by their standard deviation) for $\boldsymbol{b}$ on the line passing through $\boldsymbol{b}^*$ and $\widetilde{\boldsymbol{b}}$ for $n=3$ (a) and $n=7$ (b), where the curves indicate the contour lines, with the black curves marking
$\Lambda = 0$ and hence enclosing the region of synchronization stability (blue).
(c,d) Identified $\gamma$ values as functions of $\delta$ for $n=3$ (c) and $n=7$ (d).
(e) Minimum $b_i$-heterogeneity required for stability [corresponding to the bottom black curves in (a) and (b)] for larger networks.}
\label{fig3}
\end{center}
\end{figure}

A key property of the system allowing asymmetry-induced symmetry is the directionality of the network structure parametrized by $\delta$.
Since the difference between the link strengths in the clockwise and counterclockwise directions is $2\delta$, there is no naturally defined direction around the ring if $\delta = 0$.
When $\delta>0$, the two directions become distinguishable, indicating the absence of reflection symmetry, but the network structure remains homogeneous due to the presence of rotational symmetry.
We find that, while asymmetry-induced symmetry is not observed for $\delta = 0$, it can be observed for any $\delta > 0$, i.e., for an {\it arbitrarily small amount of this directionality}.
Indeed, for a given $\delta > 0$, we numerically identify a value of $\gamma$ for which the synchronous state is unstable at $\boldsymbol{b}^* := (b^*,\ldots,b^*)$ but stable at some $\boldsymbol b$ with heterogeneous $b_i$.
The results are shown in Fig.~\ref{fig3} for $n=3,7,15,31$, and $63$.
We see that the identified $\gamma$ values remain strictly positive, and that the stability region shrinks and moves increasingly closer to $\boldsymbol{b}^*$ as $\delta$ approaches zero.

It is interesting to interpret our results in the context of Curie's principle~\cite{Curie1894}, which asserts that the symmetries of the causes must be found in the 
effects.
Asymmetry-induced symmetry requires that 1) any state with the symmetry of the system be unstable and hence not observed and 2) the symmetry of the system be reduced to realize the symmetric state---both consistent with but not following from Curie's principle. 
For the first requirement, it must be noted that Curie's principle, which is strictly correct for exact symmetries, asserts nothing about cases involving approximate symmetries (no matter how close they are to being exact) and hence about the stability of the states~\cite{Stewart1992}.
In particular, it is not true that nearly symmetric causes lead to nearly symmetric effects, as demonstrated by the phenomenon of spontaneous symmetry breaking. This is also why the symmetric state is not observed in the system considered here, despite the symmetry of the system.
For the second requirement, while it is counterintuitive that the system has to be asymmetric in order for the symmetric state to exist and be stable, Curie's principle provides no a priori reason why an asymmetric cause could not produce a symmetric effect.
One can draw an analogy with chimera states, which are examples of symmetry breaking but not spontaneous symmetry breaking because the symmetric state is usually still stable in the system; it is not the existence of a stable state with less symmetry than the system that is striking in that case but rather the property that those states have (i.e., being stable or long lived despite being a combination of coherent and incoherent groups). 
Likewise, here too it is not the existence of a stable symmetric state for an asymmetric system that is striking but instead the fact that such state can only be stable when the system is asymmetric.

Symmetry breaking in which an asymmetric structure develops from a symmetric one plays a central role in pattern formation, of which the embryonic development of an organism has once served as a representative example. It is thus natural to ask whether the
converse, reported here for network synchronization, could have broader implications for pattern formation. We argue that it does, as it suggests a mechanism for the formation of uniform patterns out of nonuniform ones.
Examples include the development of higher-order (fivefold) radial symmetry in adult starfish from bilateral symmetry in starfish larvae~\cite{Fox:2007}, development of spherical symmetry in yeast cells from asymmetric bud cells~\cite{Slaughter:2009}, and recovery of lost symmetry in severed animals via regeneration~\cite{Abrams:2015}.
The possibility of symmetric structures developing from asymmetric ones should raise questions about the assumptions tacitly made on the causes when the effects are symmetric: while symmetry breaking allows symmetric theories to describe an observed asymmetric reality~\cite{Brading:2013}, our results show that asymmetric theories, models, or systems may be required to describe emergent symmetric patterns.

\begin{acknowledgments}
The authors thank Ferenc Molnar for help with the high-resolution visualization.
This work was supported by the Simons Foundation through Award No. 342906 and ARO Grant No. W911NF-15-1-0272.
\end{acknowledgments}


\clearpage

\onecolumngrid
\setcounter{page}{1}

\setcounter{equation}{0}
\renewcommand{\theequation}{S\arabic{equation}}

\begin{center}
{\Large\bf Supplemental Material}\\[3mm]
\textit{Symmetric States Requiring System Asymmetry}\\[1pt] 
Takashi Nishikawa and Adilson E. Motter
\end{center}

\subsection{S1.\ \ Isolated oscillator dynamics}
In the absence of coupling, each isolated oscillator in 
Eq.~\eqref{eqn:system} of the main text
belongs to the class oscillators whose governing equation
can be written in polar form as 
\begin{equation}\label{eqn:single-oscillator}
\begin{split}
\dot{\theta} &= \omega + a(r - r_c),\\
\dot{r} &= b r \left( 1 - \frac{r}{r_c} \right).
\end{split}
\end{equation}
This system has a limit cycle, which is given by $\theta = \theta_0 + \omega t$ and $r = r_c$, has constant angular frequency $\omega$ and constant amplitude $r_c>0$, and is exponentially stable with convergence rate $b > 0$.  
Note that we may assume $a = 1$ and $r_c = 1$ without loss of generality, since we can transform Eq.~\eqref{eqn:single-oscillator} into equations of the same form with $a = 1$ and $r_c = 1$ by scaling the time variable as $a r_c t \to t$ and the amplitude as $r/r_c \to r$, while redefining the other parameters as $\omega/(a r_c) \to \omega $ and $b/(a r_c) \to b$.  
In writing Eq.~\eqref{eqn:system} of the main text, 
we have assumed that the oscillators have the same value for $\omega$, $a$ ($=1$), and $r_c$ ($=1$), but can have different values for $b$.  Note that we have made both coupling terms to be proportional to $r_i$ in order to ensure that the r.h.s.\ of 
Eq.~\eqref{eqn:system} of the main text
is continuous and differentiable at $r_i = 0$, $\forall i$.
In the synchronous 
state~\eqref{eqn:sync-state} of the main text, 
each oscillator in the network follows the limit cycle mentioned above.

\subsection{S2.\ \ Stability of synchronous state}

We analyze the stability through the eigenvalues of the $2n \times 2n$ Jacobian matrix of 
system described by Eq.~\eqref{eqn:system} of the main text, 
evaluated at the synchronous state.  The Jacobian matrix can be written in a block form as
\begin{equation}\label{eqn:jacob1}
\mathbf{J} = \begin{pmatrix}
\gamma \mathbf{K} & \mathbf{1} \\
- \varepsilon\mathbf{L} & - \mathbf{D}
\end{pmatrix},
\end{equation}
where $\mathbf{K}$ is the $n \times n$ Laplacian matrix for the all-to-all coupling topology, $\mathbf{1}$ is the $n \times n$ identity matrix, $\mathbf{L}$ is the $n \times n$ Laplacian matrix corresponding to the adjacency matrix $\mathbf{A}$, and $\mathbf{D}$ is the diagonal matrix  whose diagonal components are $b_1,\ldots,b_n$.  Let us denote the eigenvalues of $\mathbf{J}$ by $\lambda_1, \lambda_2, \ldots, \lambda_{2n}$, noting that $\mathbf{J}$ always has a zero eigenvalue, $\lambda_1 = 0$, associated with eigenvector $u_1$, whose components are given by $u_{1j} = 1$ if $1 \le j \le n$ and $u_{1j} = 0$ if $n+1 \le j \le 2n$.
Since perturbations along this eigenvector do not destroy synchronization, the condition for synchronization stability is then written as 
\begin{equation}
\Lambda := \max_{2 \le j \le 2n} \text{Re}(\lambda_j) < 0.
\end{equation}
Here $\Lambda$ is the maximum Lyapunov exponent, which measures the exponential rate of convergence to (or divergence from, if $\Lambda > 0$) the synchronous state,
thus providing a quantitative measure of the strength of synchronization stability.

\subsection{S3.\ \ Networks of coupled Stuart-Landau oscillators}

The Stuart-Landau equation is derived from the normal form of a supercritical Hopf bifurcation~\cite{kuramoto2003chemical} and takes the following general form:
\begin{equation}\label{eqn:SL1}
\dot{z} = c_1 z - c_3 z |z|^2,
\end{equation}
where $z$ is a complex variable, and $c_1$ and $c_3$ are complex coefficients.
Assuming that a limit cycle exists, normalizing $z$ by the amplitude of the limit cycle, and re-parameterizing the coefficients in Eq.~\eqref{eqn:SL1}, we obtain
\begin{equation}\label{eqn:SL2}
\dot{z} = [b + \mathbf{i} (\omega + a)] z - (b + \mathbf{i} a) z |z|^2,
\end{equation}
where $b>0$ follows from the existence of the limit cycle (thus making Eq.~\eqref{eqn:SL2} correspond to the post-bifurcation regime of this supercritical Hopf bifurcation) and we denote the imaginary unit as $\mathbf{i} := \sqrt{-1}$.
The limit cycle follows the unit circle in the complex plane with constant angular frequency $\omega$, and hence is given by $z(t) = e^{ \mathbf{i} (\omega t + \theta_0)}$ for some constant $\theta_0$.
The constant $a$ parametrizes the dependence of the angular frequency on the amplitude $|z|$. 
This can be seen when writing Eq.~\eqref{eqn:SL2} in polar form using $z = r e^{\mathbf{i} \theta}$:
\begin{equation}\label{eqn:SL3}
\begin{split}
\dot{\theta} &=  \omega + a(1 - r^2 ),\\
\dot{r} &= b r (1 - r^2).
\end{split}
\end{equation}
In this form the limit cycle is given by $\theta(t) = \omega t + \theta_0$, $r(t) = 1$.
Equations~\eqref{eqn:SL2} and \eqref{eqn:SL3} are thus equivalent descriptions of the dynamics of a single Stuart-Landau oscillator.
While Eq.~\eqref{eqn:SL3} is different from Eq.~\eqref{eqn:single-oscillator}, the role of parameter $b$ is the same because in both cases $b$ gives the exponential rate of convergence to the limit cycle.

We consider a network of $n$ diffusively coupled Stuart-Landau oscillators whose dynamics is governed by
\begin{equation}\label{eqn:coupled-SL1}
\dot{z_i} = [b_i + \mathbf{i} (\omega + a)] z_i - (b_i + \mathbf{i} a) z_i |z_i|^2 + \sum_{j=1}^n C_{ij} \left(\frac{z_j}{|z_j|} - \frac{z_i}{|z_i|} \right).
\end{equation}
Note that the coupling is through $z_i/|z_i| = e^{\mathbf{i}\theta_i}$, which depends only on the angle variable $\theta_i$.
The matrix $\mathbf{C} := (C_{ij})_{1 \le i,j \le n}$ of coupling coefficients can be interpreted as a complex-valued adjacency matrix of the network.
Defining the corresponding Laplacian matrix $\mathbf{G}$ by $G_{ij} = - C_{ij}$ if $i \neq j$ and $G_{ii} = \sum_{k \neq i} C_{ik}$, the coupling term in Eq.~\eqref{eqn:coupled-SL1} can also be written as $- \sum_{j=1}^n G_{ij} z_j/|z_j|$.
Note that we have assumed that the limit cycle frequency $\omega$ and the parameter $a$ are identical for all oscillators, while the parameter $b_i$ can be different for different oscillators.
System~\eqref{eqn:coupled-SL1} has a synchronous state (which can be stable or unstable) given by $z_i(t) = e^{ \mathbf{i} (\omega t + \theta_0)}$ for all $i$.
In polar form, Eq.~\eqref{eqn:coupled-SL1} can be written as
\begin{equation}\label{eqn:coupled-SL2}
\begin{split}
\dot{\theta_i} &=  \omega + a(1 - r_i^2 ) + \frac{1}{r_i} \sum_{j=1}^n \text{Re}(C_{ij}) \sin(\theta_j - \theta_i) + \frac{1}{r_i} \sum_{j=1}^n \text{Im}(C_{ij}) [\cos(\theta_j - \theta_i) - 1],\\
\dot{r_i} &= b_i r_i (1 - r_i^2) + \sum_{j=1}^n \text{Re}(C_{ij}) [\cos(\theta_j - \theta_i) - 1] - \sum_{j=1}^n \text{Im}(C_{ij}) \sin(\theta_j - \theta_i).
\end{split}
\end{equation}
It can be shown that the Jacobian matrix of this system, evaluated at the synchronous state, can be written using the Laplacian matrix $\mathbf{G}$ as
\begin{equation}\label{eqn:jacob2}
\begin{pmatrix}
-\text{Re}(\mathbf{G}) & -2a\mathbf{1} \\
\text{Im}(\mathbf{G}) & -2\mathbf{D}
\end{pmatrix},
\end{equation}
where we recall that $\mathbf{1}$ denotes the $n \times n$ identity matrix and $\mathbf{D}$ is the diagonal matrix whose diagonal components are $b_1,\ldots,b_n$.
We see that this Jacobian matrix becomes identical to the one in Eq.~\eqref{eqn:jacob1} if we set $a = -1/2$, scale the parameters $b_i$ as $b_i \to b_i/2$, and let $\text{Re}(\mathbf{C}) = -\gamma \mathbf{A}'$, $\text{Im}(\mathbf{C}) = - \varepsilon \mathbf{A}$, where $\mathbf{A}'$ is the adjacency matrix of the (unweighted) all-to-all network (i.e., $A'_{ij} = 1$, $\forall i \neq j$) and $\mathbf{A}$ is the same adjacency matrix used for Eq.~\eqref{eqn:jacob1}.
Since the maximum Lyapunov exponent is the maximum real part of the eigenvalue of the Jacobian matrix in Eq.~\eqref{eqn:jacob2}, we see that the $b_i$-dependence of the stability of the synchronous state is identical to system (1) of the main text.

\subsection{S4.\ \ Simulation of network dynamics}

To simulate the dynamics for 
Fig.~\ref{fig1} of the main text
we integrated 
Eq.~\eqref{eqn:system}
using $\delta = 0.3$, $\gamma = 0.1$, $\varepsilon = 2$, and $\omega = 1$.
For $0 \le t < 75$ we set $b_i$ as indicated by the red numbers in 
Fig.~\ref{fig1}(a).
The common value $b^* \approx 1.868$ for $b_i$ was computed by numerically minimizing $\Lambda$ under the constraint of equal $b_i$, using Matlab's implementation of the simplex algorithm in Ref.~\cite{Lagarias1998}.
For $75 \le t \le 150$ we set $b_i$ as indicated by the blue numbers in 
Fig.~\ref{fig1}(a), 
which was found by the numerical minimization of $\Lambda$ without the equal $b_i$ constraint but with the constraint $b_i > 0$, $\forall i$.  To solve this constrained optimization problem we used  Matlab's implementation of the interior point algorithm~\cite{Byrd1999}.

\subsection{S5.\ \ Animated demonstration of asymmetry-induced symmetry}

\href{https://youtu.be/zP_6EuMzt1I}{Supplemental Movie} (\url{https://youtu.be/zP_6EuMzt1I}) shows the dynamics of the
example network in Fig.~1 of the main text, in which oscillator heterogeneity is required to stabilize a homogeneous, synchronous state.
In panel~(a), circles represent the limit cycle each oscillator would follow in the absence of coupling. The $b_i$ value of the $i$th oscillator is shown inside the corresponding circle.  Starting nearly synchronized, the oscillators with homogeneous $b_i$ desynchronize and approach a traveling-wave state; after making $b_i$ values suitably heterogeneous at $t = 75$, the oscillators converge spontaneously to the synchronous state, indicating that the state is now stable.  
Panels~(b)--(f) show an animation of Fig.~1(b)--(f) of the main text, in which moving colored dots are used to visualize the dynamics of the individual oscillators.

\begin{center}
\includegraphics[width=7in]{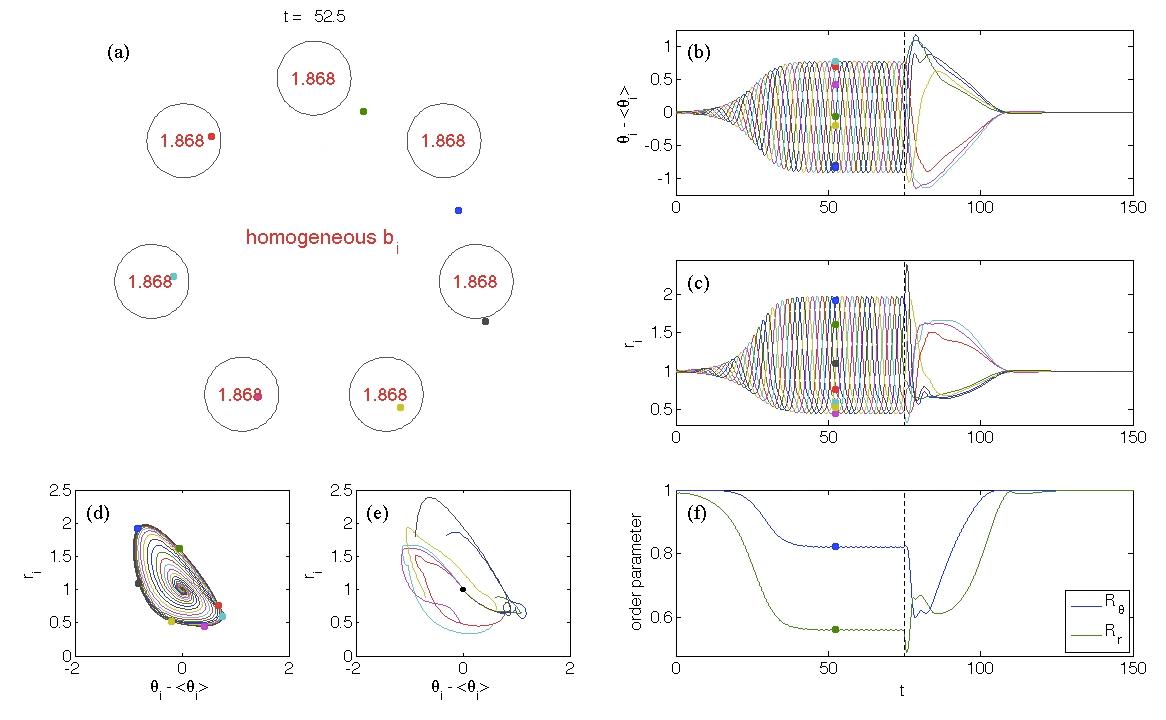}
\mbox{FIG.~S1. Snapshot from \href{https://youtu.be/zP_6EuMzt1I}{Supplemental Movie} demonstrating asymmetry-induced symmetry.}
\end{center}

\clearpage

\subsection{S6.\ \ Nonuniform networks requiring oscillator heterogeneity for synchronization stability}

To demonstrate that there are nonuniform networks for which nonuniform $b_i$ values are necessary for the stability of the (uniform) synchronous state, we have generated random undirected unweighted networks of size $n = 5$ with $4$, $6$, and $8$ links.
Figure~S2 below shows, for each number of links, one representative network that has a $\gamma$ value for which (1) the maximum Lyapunov exponent $\Lambda>0$ for all possible uniform $b_i$ (i.e., for all $b$ with $b_1 = \cdots = b_5 = b$), and (2) there are nonuniform $b_i$ values for which $\Lambda<0$.  

\begin{center}
\includegraphics{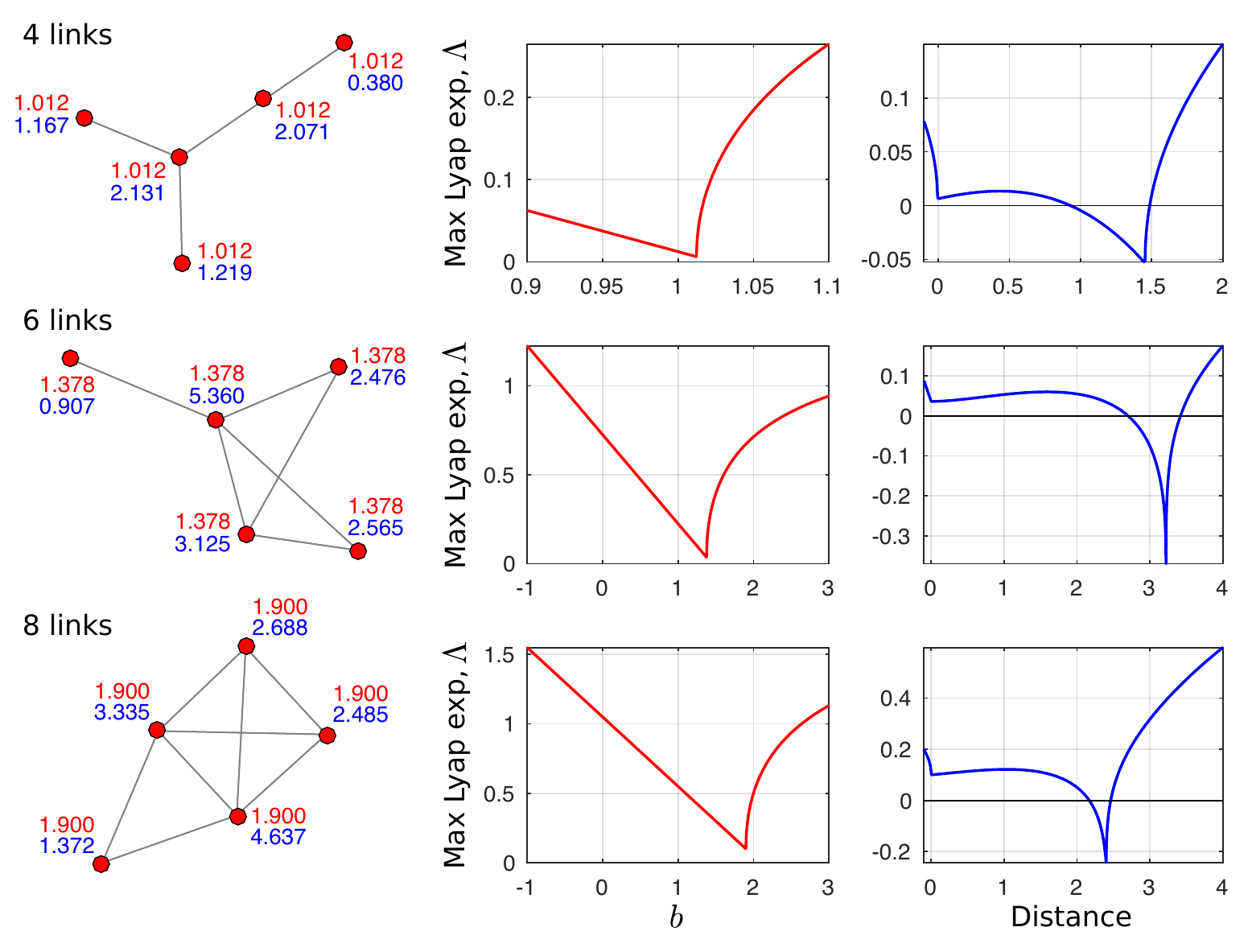}\\
\end{center}
FIG.~S2. Maximum Lyapunov exponent $\Lambda$ vs.\ parameter $b$ with $b_i = b$ for all $i$ (middle column) and vs.\ the distance from the best uniform $b_i$ values in the five-dimensional $b_i$-space  (right column) for three examples of nonuniform networks (left column).
The nodes are labeled with the optimal uniform (red) and nonuniform (blue) $b_i$ values.
In these examples we used $\delta = 0.3$, $\varepsilon = 2$, and $\gamma=0.205$ (top row), $\gamma=0.290$ (middle row), or $\gamma=0.420$ (bottom row).

\subsection{S7.\ \ Scaling property of $\boldsymbol{\Lambda}$}

Let us write $\mathbf{J} = \mathbf{J}(\varepsilon, \gamma, \mathbf{D}, \mathbf{L})$ and define $\mathbf{D'}$ to be the diagonal matrix with diagonal elements $b'_1,\ldots,b'_n$.
The characteristic polynomial of $\mathbf{J}$ can be written as
\begin{equation}
\begin{split}
\det&(\mathbf{J} - \lambda \mathbf{1})
= \det\bigl((-\gamma \mathbf{K}\mathbf{D} + \varepsilon\mathbf{L}) 
+ \lambda(\mathbf{D} - \gamma\mathbf{K}) + \lambda^2\mathbf{1}\bigr)\\
&= \det\Bigl(\textstyle\frac{\varepsilon}{\varepsilon_0}(-\gamma'\mathbf{K}\mathbf{D}' + \varepsilon_0\mathbf{L}) 
+ \textstyle\frac{\varepsilon}{\varepsilon_0}\lambda'(\mathbf{D}' - \gamma'\mathbf{K})
+ \textstyle\frac{\varepsilon}{\varepsilon_0}(\lambda')^2\mathbf{1} \Bigr)\\
&= \bigl(\textstyle\frac{\varepsilon}{\varepsilon_0}\bigr)^n
\det\Bigl((-\gamma'\mathbf{K}\mathbf{D}' + \varepsilon_0\mathbf{L}) 
+ \lambda'(\mathbf{D}' - \gamma'\mathbf{K})
+ (\lambda')^2\mathbf{1} \Bigr),
\end{split}
\end{equation}
where $\lambda' := \lambda\sqrt{\varepsilon_0/\varepsilon}$.
Notice that the determinant in the last expression is precisely the characteristic polynomial of $\mathbf{J}(\varepsilon_0, \gamma', \mathbf{D}', \mathbf{L})$ with $\lambda$ replaced by $\lambda'$.
Thus, we have that $\lambda$ is an eigenvalue of $\mathbf{J}(\varepsilon, \gamma, \mathbf{D}, \mathbf{L})$ if and only if $\lambda'$ is an eigenvalue of $\mathbf{J}(\varepsilon_0, \gamma', \mathbf{D}', \mathbf{L})$.  This immediately leads to the scaling result in 
Eq.~(3) of the main text.

\subsection{S8.\ \ Verifying asymmetry-induced symmetry for any $\boldsymbol{\delta>0}$}

We formulate a bi-layer optimization problem to identify a value of $\gamma$ and a direction $\Delta\boldsymbol{b}$ for each $\delta > 0$, so that $\Lambda(\boldsymbol{b}^*) > 0$ and $\Lambda(\boldsymbol{b}^* + s \Delta\boldsymbol{b}) < 0$ for some $s$.
For a given $\delta > 0$, we define an objective function
\begin{equation}
F(\gamma) := \frac{1}{2}\Bigl\vert \Lambda(\boldsymbol{b}^*; \gamma, \delta)
+ \min_{0 \le s \le 10} \Lambda(\boldsymbol{b}^* + s \Delta\boldsymbol{b}; \gamma, \delta) \Bigr\vert,
\end{equation}
where $\Delta\boldsymbol{b}$ is the normalized eigenvector corresponding to the smallest eigenvalue of the Hessian matrix of the function $\Lambda(\boldsymbol{b})$ at $\boldsymbol{b}^*$.
Note that $\boldsymbol{b}^*$ and $\Delta\boldsymbol{b}$ generally depend on $\gamma$ and that our choice of $\Delta\boldsymbol{b}$ ensures that the quadratic decrease of $\Lambda$ near $\boldsymbol{b}^*$ along this direction is the fastest possible.
By minimizing $F(\gamma)$ on the interval $[0,10]$, we seek to make the average between the $\Lambda$ value at $\boldsymbol{b}^*$ and the minimum $\Lambda$ value at $\boldsymbol{b}^* + s \Delta\boldsymbol{b}$ as close to zero as possible.
This allows us to search for a $\gamma$ value for which the two $\Lambda$ values straddle zero.
The two levels of minimization involved in this procedure were solved using Matlab's \texttt{fminbnd} function, which is based on golden section search and parabolic interpolation~\cite{Forsythe1976}.
Using the resulting value of $s$, we set $\widetilde{\boldsymbol{b}}:=\boldsymbol{b}^* + s \Delta\boldsymbol{b}$.
For the $n=3$ case in 
Fig.~\ref{fig3} of the main text,
the choice of $\Delta\boldsymbol{b}$ is not unique,
since the smallest eigenvalue of the Hessian matrix is doubly degenerate.
We thus made a specific choice within the two-dimensional eigenspace: $\Delta\boldsymbol{b} = (1,1,-2)^T/\sqrt{6}$.
For the $n=63$ case, we first compute $\Delta\boldsymbol{b}$ for $\gamma=0$ and use this fixed direction as an approximation to $\Delta\boldsymbol{b}$ for all $\gamma$ values to make this case computationally tractable.


\begin{thebibliography}{10}

\bibitem{Weyl1952}
H. Weyl,
{\it Symmetry}
(Princeton University Press, Princeton, NJ, 1952).

\bibitem{Anderson1972}
P.~W. Anderson, Science {\bf 177}, 393 (1972).

\bibitem{MacArthur2008}
B.~D. MacArthur, R.~J. S\'anchez-Garc\'{i}a, and J.~W. Anderson,  Discrete Appl. Math. {\bf 156}, 3525 (2008).

\bibitem{Pecora2014}
L.~M. Pecora, F. Sorrentino, A.~M. Hagerstrom, T.~E. Murphy, and R. Roy,  Nat. Commun. {\bf 5}, 4079 (2014).

\bibitem{Kuramoto2002}
Y. Kuramoto and D. Battogtokh, Nonl. Phen. Compl. Syst. {\bf 5}, 380 (2002).

\bibitem{Abrams2004}
D.~M. Abrams and S.~H. Strogatz, Phys. Rev. Lett. {\bf 93}, 174102 (2004).

\bibitem{Ott2008}
E. Ott and T.~M. Antonsen, Chaos {\bf 18}, 037113 (2008).

\bibitem{Motter2010}
A.~E. Motter, Nat. Phys. {\bf 6}, 164 (2010).

\bibitem{Pikovsky2003}
A. Pikovsky, M. Rosenblum, and J. Kurths, {\it Synchronization: A Universal Concept in Nonlinear Sciences} (Cambridge University Press, Cambridge, England, 2003).

\bibitem{4700861}
P. Yang, R. Freeman, and K. Lynch, IEEE Trans. Automat. Contr. {\bf 53}, 2480 (2008).

\bibitem{4140748}
W. Ren and R.~W. Beard, {\it Distributed Consensus in Multi-vehicle Cooperative Control} (Springer-Verlag, Berlin, 2008).

\bibitem{DeGroot:1974}
M.~H. DeGroot, J. Am. Stat. Assoc. {\bf 69}, 118 (1974).

\bibitem{Friedkin:2011}
N.~E. Friedkin and E.~C. Johnsen, {\it Social Influence Network Theory: A Sociological Examination of Small Group Dynamics} (Cambridge University Press, Cambridge, England, 2011).

\bibitem{Showalter:2015}
K. Showalter and I.~R. Epstein, Chaos {\bf 25}, 097613 (2015).

\bibitem{Tompkins:2015}
N. Tompkins, M.~C. Cambria, A.~L. Wang, M. Heymann, and S. Fraden, Creation and perturbation of planar networks of chemical oscillators.
{\it Chaos} {\bf 25}, 064611 (2015).

\bibitem{Nakao:2010fk}
H. Nakao and A.~S. Mikhailov, Nat. Phys. {\bf 6}, 544 (2010).

\bibitem{maas1987transportation}
C. Maas, Discrete Appl. Math. {\bf 16}, 31 (1987).

\bibitem{methods}
See Supplemental Material for details on the isolated oscillator dynamics (Sec.~S1), stability of the synchronous state (Sec.~S2), networks of coupled Stuart-Landau oscillators (Sec.~S3), simulation of network dynamics (Sec.~S4), animated version of Fig.~\ref{fig1} (Sec.~S5 and an associated movie at \url{https://youtu.be/zP_6EuMzt1I}), nonuniform networks requiring oscillator heterogeneity for synchronization stability (Sec.~S6), scaling property of $\Lambda$ in Eq.~\eqref{eqn:lambda-max-scaling} (Sec.~S7), and demonstration of asymmetry-induced symmetry for any $\delta>0$ (Sec.~S8).

\bibitem{Daido:2004}
H. Daido and K. Nakanishi, Phys. Rev. Lett. {\bf 93}, 104101 (2004).

\bibitem{Bordyugov:2010}
G. Bordyugov, A. Pikovsky, and M. Rosenblum, Phys. Rev. E {\bf 82}, 035205(R) (2010).

\bibitem{Laing:2010}
C.~R. Laing, Phys. Rev. E {\bf 81}, 066221 (2010).

\bibitem{Zakharova:2014}
A. Zakharova, M. Kapeller, and E. Sch\"oll, Phys. Rev. Lett. {\bf 112}, 154101 (2014).

\bibitem{Panaggio:2015}
M.~J. Panaggio and D.~M. Abrams, Nonlinearity {\bf 28}, R67 (2015).

\bibitem{kuramoto2003chemical}
Y.~Kuramoto,
{\it Chemical Oscillations, Waves, and Turbulence}
(Dover Publications, New York, 2003).

\bibitem{Curie1894}
P. Curie, J. Phys. Th\'eor. Appl. {\bf 3}, 393 (1894).

\bibitem{Stewart1992}
I. Stewart and M. Golubitsky, {\it Fearful Symmetry: Is God a Geometer?}
(Blackwell, Cambridge, MA, 1992).

\bibitem{Fox:2007}
R. Fox, {\it Invertebrate Zoology OnLine}, \url{http://lanwebs.lander.edu/faculty/rsfox/invertebrates/}.

\bibitem{Slaughter:2009}
B.~D. Slaughter, S.~E. Smith, and R. Li, Cold Spring Harbor Perspect. Biol. {\bf 1}, a003384 (2009).

\bibitem{Abrams:2015}
M.~J. Abrams, T. Basinger, W. Yuan, C.-L. Guo, and L. Goentoro, Proc. Natl. Acad. Sci. U.S.A. {\bf 112}, E3365 (2015).

\bibitem{Brading:2013}
K. Brading and E. Castellani, ``Symmetry and symmetry breaking'' in {\it The Stanford Encyclopedia of Philosophy}, edited by E.~N. Zalta (Stanford University, Standford, CA, 2013).

\bibitem{Lagarias1998}
J.~C. Lagarias, J.~A. Reeds, M.~H. Wright, and P.~E. Wright, SIAM J. Optimiz. {\bf 9}, 112 (1998).

\bibitem{Byrd1999}
R.~H. Byrd, M.~E. Hribar, and J. Nocedal, SIAM J. Optimiz. {\bf 9}, 877 (1999).

\bibitem{Forsythe1976}
G.~E. Forsythe, M.~A. Malcolm, and C.~B. Moler, {\it Computer Methods for Mathematical Computations} (Prentice-Hall, 1976).

\end{thebibliography}
\end{document}